\documentclass[showkeywords,preprint,amssymb,amsmath,natbib,%
   superscriptaddress]{revtex4}
\usepackage{amssymb}
\usepackage{epsfig}

\begin{document}

\title{ Thermalized non-equilibrated matter and high temperature
superconducting state in quantum many-body systems}

\author{L. Benet}%
\affiliation{Instituto de Ciencias F\'{\i}sicas,
  Univeridad Nacional Aut\'onoma de M\'exico (UNAM)\\
  Apdo. Postal 48--3, 62251--Cuernavaca, Mor., M\'exico}%
\author{M. Bienert}%
\affiliation{Instituto de Ciencias F\'{\i}sicas,
  Univeridad Nacional Aut\'onoma de M\'exico (UNAM)\\
  Apdo. Postal 48--3, 62251--Cuernavaca, Mor., M\'exico}%
\author{S.Yu. Kun}%
\affiliation{Facultad de Ciencias, Universidad del Estado de Morelos (UAEM)\\
  62209-Cuernavaca, Morelos, Mexico}%
\affiliation{Centre for Nonlinear Physics, RSPhysSE,
  Australian National University\\
  Canberra ACT 0200, Australia}%

\date{\today}

\begin{abstract}
  A characteristic feature of thermalized non-equilibrated matter is
  that, in spite of energy relaxation--equilibration, a phase memory
  of the way the many-body system was excited remains. As an example,
  we analyze data on a strong forward peaking of thermal proton yield
  in the Bi($\gamma$,p) photonuclear reaction. New analysis shows that
  the phase relaxation in highly-excited heavy nuclei can be 8 orders
  of magnitude or even much longer than the energy relaxation. We
  argue that thermalized non-equilibrated matter resembles a
  high temperature superconducting state in quantum many-body
  systems. We briefly present results on the time-dependent
  correlation function of the many-particle density fluctuations for
  such a superconducting state. It should be of interest to
  experimentally search for manifestations of thermalized
  non-equilibrated matter in many-body mesoscopic systems and
  nanostructures.\\
  {~}
\end{abstract}


\maketitle


\section{Introduction}
\label{sec1}

Consider a beam of photons or electrons directed on a many-electron
quantum dot. Let the quantum dot be three dimensional with spherically
symmetric or two dimensional with circularly symmetric confining
potential. Upon the interaction of the incoming radiation with the
electrons inside the dot and because of the inter-dot
electron--electron interaction the total energy of the system
eventually gets redistributed among many electrons: The quasi-bound
many-electron quantum dot reaches a thermally equilibrated
state. Then, after a certain period of time, due to the strong
electron--electron interaction, a sufficient energy may be
concentrated on a single electron for its escape from the dot. Such a
thermal emission can also be viewed as an evaporation process usually
described by the statistical reaction theory or phase space
theory~\cite{campb}.

We ask the question: Do angular distributions for the thermal emission
carry any information of the way the quantum dot was excited? Is there
any memory about a direction of the beam of incoming radiation for the
thermal emission? More specifically, are angular distributions of the
thermal emission necessarily symmetric about 90$^\circ$ in the center
of mass system (c.m.) with respect to the direction of the incident
beam? These questions have never been addressed
experimentally. Indeed, in modern physics the notions
``thermalization'' and ``energy equilibration'' are considered to be
equivalent to the notion ``statistical equilibrium''. This equivalence
seems so obvious that it has never been questioned for highly-excited
quantum many-body systems.

While for the above stated reasons the problem has never been
addressed either theoretically or experimentally for mesoscopic
systems, {\sl e.g.} for many-electron quantum dots, it turns out that
there are many well--documented nuclear data sets which reveal
unexpected and counterintuitive forward peaking for a thermal emission
from highly excited quantum many-body systems. Curiously, some of
these data sets have been available for longer than 50 years. Yet, the
fact that thermal emission from a compound nucleus can demonstrate a
strong angular asymmetry around 90$^\circ$ c.m. has never been
recognized by nuclear physicists. Accordingly, the effect has been
unknown to a wide physics community. As an illustration, we present an
extended analysis of the data on a strong forward peaking of thermal
proton yield in the Bi($\gamma$,p) photonuclear reaction. The effect
is described in terms of anomalously slow phase relaxation in
highly-excited quantum many-body systems. This effect is of
significant implications for multi-qubit ($n\simeq 100-1000$) quantum
computers since it can extend the time for quantum computing far
beyond the quantum chaos border~\cite{flores,sigma}. The effect of
anomalously slow phase relaxation has also been revealed for heavy ion
collisions~\cite{kun97HI,KunVagVor99,KunRobVag,KunPRL00,KVG01,KunChVG02,%
  KunBChGrH03,BKWD05,BenetKW05} and bimolecular chemical
reactions~\cite{BenetKW05,BCKW06}.  We find that the phase memory in
highly-excited heavy nuclei can be 8 orders of magnitude or even much
longer than the energy relaxation. We argue that a new form of matter
- thermalized non-equilibrated matter, introduced by one of
us~\cite{kun:94,kun:97b}, resembles a high temperature superconducting
state in quantum many-body systems. It should be of interest to
experimentally search for manifestations of thermalized
non-equilibrated matter in many-body mesoscopic systems and
nanostructures.

In the Appendix, we briefly outline the results on time-dependent
correlation function of the many-particle density fluctuations in such
a high temperature superconducting state of the thermalized
non-equilibrated matter.

\section{Experimental evidence for a formation of thermalized
  non-equilibrated matter in ${\rm Bi}$($\gamma, p$) photonuclear
  proton evaporation}
\label{sec2}

We analyze the proton yield of the Bi($\gamma$,p) photonuclear
reaction produced by 24~MeV bremsstrahlung. The properly scaled
angle-integrated spectrum for the proton energy $\varepsilon \leq
8$~MeV has an exponential shape with a slope of
0.55~MeV~\cite{sigma}. This is characteristic for the decay of
thermalized compound nucleus with a ``temperature'' $T=0.55$~MeV of
the residual nucleus. The average excitation energy of the compound
nucleus can be evaluated as $\bar E^\ast =14$~MeV, {\sl i.e.} slightly
above the center of the dipole giant resonance peak at
13.5~MeV~\cite{Bigiantres64}. Then, the experimentally determined
nuclear ``temperature'' is in a good agreement with the statistical
model calculations~\cite{sigma}.

\begin{figure}
  \centerline{\includegraphics[angle=0,width=12cm]{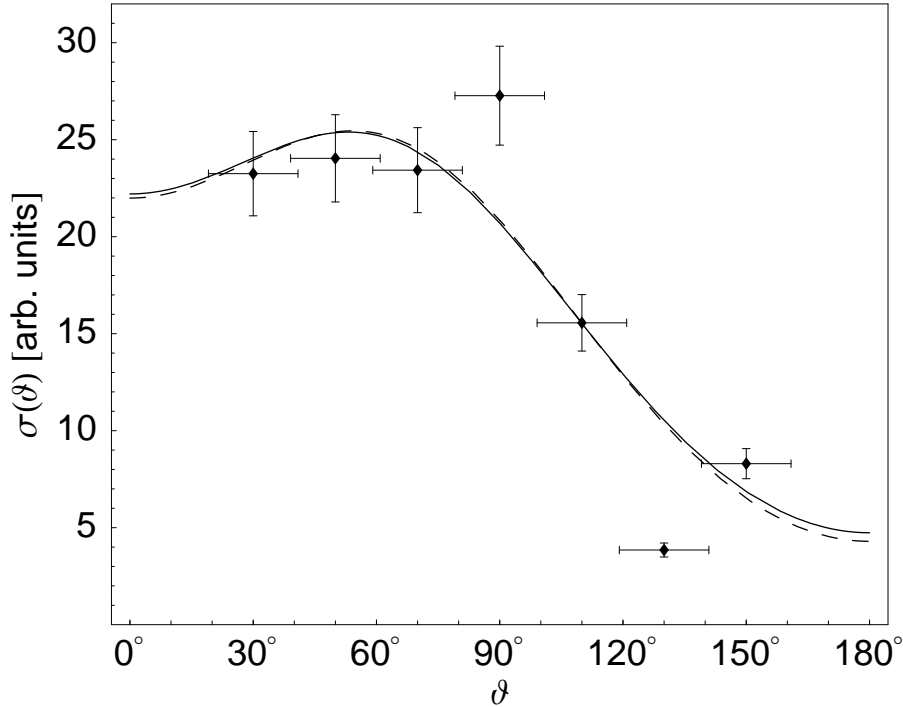}}%
  \caption{ \label{fig1}%
   Experimental proton angular distributions (in arbitrary units) from
   the Bi($\gamma$,p) photonuclear evaporation process for
   $\varepsilon =2-8$~MeV~\cite{dataBi}. The solid curve is a fit to
   the experimental data with $\beta/\Gamma_{cn}=0.11$, and the dashed
   curve is a fit with $\beta/\Gamma_{cn}\to 0$, (see text).  }
\end{figure}

In Fig.~\ref{fig1} we present experimental proton angular
distributions from the Bi($\gamma$,p) photonuclear reaction for
$\varepsilon =2-8$~MeV~\cite{dataBi}. We observe that, in spite of
complete energy relaxation in the thermalized compound nucleus, the
angular distributions are strongly asymmetric about 90$^\circ $, {\sl
i.e.}, memory of the direction of the incident $\gamma$-ray beam is
clearly retained. Therefore, even though the compound nucleus is in a
thermalized state, it is far from being fully equilibrated.

\section{Theoretical interpretation of the forward peaking of
evaporating protons in the ${\rm Bi}$($\gamma, p$) 
photonuclear reaction}
\label{sec3}

Clearly, a description of the decay of thermalized but yet
non-equilibrated matter requires a major modification of conventional
picture of compound nucleus (see {\it e.g.}  Ref.~\cite{blatt})
originally formulated by Bohr, Bethe, Weisskopf, Wigner, Dyson and
others. The basic idea behind the conventional picture is that
thermalization of the compound nucleus guarantees a complete loss of
memory of initial phase relations. A modification of this conventional
picture of the compound nucleus was proposed by one of us in
Refs.~\cite{kun:94,kun:97b}. Unfortunately, there has been no other
interpretations of an angular asymmetry around 90$^\circ$ c.m. in
evaporation processes. The key element in the description of the
asymmetry of angular distributions around 90$^\circ$ c.m. for
evaporating particles is the total spin off-diagonal correlation
between compound nucleus partial width amplitudes. Such correlation is
neglected in a conventional picture of compound
nucleus. Following~\cite{kun:94,kun:97b}, we have
\begin{equation}
\label{eq1}
\frac{\overline{\gamma_{\mu_1}^{J_1a_1}
\gamma_{\mu_1}^{J_1b_1}\gamma_{\mu_2}^{J_2a_2}
\gamma_{\mu_2}^{J_2b_2}}}
{\Big[\overline{(\gamma_{\mu_1}^{J_1a_1})^2}\,
       \overline{(\gamma_{\mu_1}^{J_1b_1})^2}\,
       \overline{(\gamma_{\mu_2}^{J_2a_2})^2}\,
       \overline{(\gamma_{\mu_2}^{J_2b_2})^2}\,\Big]^{1/2}} = 
\frac{(1/{\pi}) D \beta |J_1-J_2|}{
(E_{\mu_1}^{J_1}-E_{\mu_2}^{J_2})^2+ \beta^2(J_1-J_2)^2} \, ,
\end{equation}
where the overlines denote ensemble averaging.  Here, $J_1\neq J_2$
are the compound nucleus total spin values, $E_\mu^{J}$ are the
resonance energies with $\mu$ being running indices, $D$ is average
level spacing of the compound nucleus, and $\gamma_\mu^{J a}$ are the
partial width amplitudes for the formation and decay of the compound
nucleus. The $a(b)$ indices specify the orbital momenta
$l_{a_{1,2}}(l_{b_{1,2}})$, the channel spins
$j_{a_{1,2}}(j_{b_{1,2}})$, and the microstates $\bar a (\bar b )$ of
the target and residual nucleus, respectively. Accordingly, $\bar
a_1=\bar a_2 $ denotes the ground state of the target, and $\bar
b_1=\bar b_2 $ specifies the microstates of the residual nucleus.  The
phase relaxation width $\beta$, introduced in
Refs.~\cite{kun:93,kun:94,kun:97b}, determines a characteristic time,
$\tau_{ph}=\hbar/\beta$, for the decay of the spin off-diagonal phase
correlations. The above correlation between the partial width
amplitudes leads to a correlation between fluctuating compound nucleus
$S$-matrix elements carrying different total spin values,
\begin{equation}
\label{eq2}
\langle S_{a_1b_1}^{J_1}(E)^\ast\, S_{a_2b_2}^{J_2}(E) \rangle
= \frac{
\Big[ \langle |S_{a_1b_1}^{J_1}(E)|^2\rangle 
      \langle |S_{a_2b_2}^{J_2}(E)|^2\rangle\Big]^{1/2}}
{1+|J_1-J_2|\beta / \Gamma_{cn}}.
\end{equation}
Here, $\Gamma_{cn}$ is the compound nucleus decay width,
$S_{ab}^{J}(E)$ are compound nucleus $S$-matrix elements with total
spin $J$ and the brackets $\langle...\rangle$ denote the energy $E$
averaging. For finite values of $\beta / \Gamma_{cn}$, non-vanishing
spin off-diagonal correlations in Eq. (2) reflect non-vanishing
interference between resonance levels with different total spins upon
energy averaging.

For the correlation between $S$-matrix elements carrying the same
total spin values and the same microstates $\bar a_1=\bar
a_2$ and $\bar b_1=\bar b_2$, but different orbital momenta and/or
channel spins, we have~\cite{kun:94,kun:97b}
\begin{equation}
\label{eq3}
\langle S_{a_1b_1}^{J}(E)^\ast S_{a_2b_2}^{J}(E) \rangle =
[<|S_{a_1b_1}^{J}(E)|^2> <|S_{a_2b_2}^{J}(E)|^2>]^{1/2}.
\end{equation}
The above equation results from a strong correlation between
the partial width amplitudes $\gamma_\mu^{J a_1(b_1)}$ and
$\gamma_\mu^{J a_2(b_2)}$ with $\bar a_1=\bar a_2$ and $\bar
b_1=\bar b_2$ but $l_{a_1}\neq l_{a_2}$, $l_{b_1}\neq l_{b_2}$,
$j_{a_1}\neq j_{a_2}$, $j_{b_1}\neq j_{b_2}$. Such a correlation is
referred to~\cite{kun:94,kun:97b} as the continuum correlation.

For $\beta \gg \Gamma_{cn}$, the spin off-diagonal correlations in
Eq.~(\ref{eq2}) result in the angular distributions symmetric around
90$^\circ$ c.m., recovering the conventional Bohr picture of the
compound nucleus. However, if $\beta\sim \Gamma_{cn}$, {\sl i.e.} the
phase relaxation time $\tau_{ph}=\hbar/\beta$ is comparable or longer
than the average life--time of the compound nucleus
$\hbar/\Gamma_{cn}$, this allows us to describe a strong asymmetry of
the angular distributions of the evaporating yield around 90$^\circ$
c.m.

We analyze the angular distribution of the thermal proton yield in
Fig. 1 following Ref.~\cite{sigma}. Without repeating details we note
that the shape of the angular distribution has been found to depend on
the four parameters: $A=T^{L=2}/T^{L=1}$, $B=T^{l^\prime
  =1}/T^{l^\prime =0}$, $C=T^{l^\prime =2}/T^{l^\prime =0}$, and
$\beta/\Gamma_{cn}$. Here, $T^L$ are the entrance channel transmission
coefficients for the formation of the compound nucleus with the total
spins $L=1$ and $L=2$ due to the absorption of electric dipole and
quadrupole radiation, accordingly.  The exit channel transmission
coefficients $T^{l^\prime }$ with $l^\prime =0,1,2$ being orbital
momenta of the evaporated protons have been assumed to be independent
of the compound nucleus spin $L$ and the spin of the residual
nucleus~\cite{blatt}.

From the best fit of the experimental angular distributions in Fig. 1
we obtain: $A=0.082$, $B=0.47$, $C=0.37$ and $\beta/\Gamma_{cn}=0.11$.
The compound nucleus decay width $\Gamma_{cn}$ for Bi with an
excitation energy of 14~MeV can be estimated from the systematics in
Fig. 7 of Ref.~\cite{ericson}, which provides a good description of
the experimentally determined $\Gamma_{cn}$ for a wide range of mass
numbers. From this estimation we obtain $\Gamma_{cn}\simeq 0.1$~eV
yielding $\beta\simeq 0.01$~eV. At the same time, the standard nuclear
physics estimate for the spreading width of the Bi nucleus with the
excitation energy 14~MeV is about 2~MeV (see Fig.~2.1 in
Ref.~\cite{AWM75}). This is close to another estimate of
$\Gamma_{spr}$ as the width of a dipole giant
resonance~\cite{anderson}, which is about 4.5~MeV for
Bi~\cite{Bigiantres64}.  Notice that $\hbar/\Gamma_{spr}$ is the
energy relaxation time and $\hbar/\beta $ is the phase relaxation
time. Therefore we observe that the phase relaxation is at least 8
orders of magnitude slower than energy relaxation.

In Fig.~\ref{fig1} we also present the best fit for $\beta/\Gamma_{\rm
cn}\to 0$ (but still with $D/\beta \ll 1$~\cite{kun:97b}). It is
obtained with $A=0.066$, $B=0.42$ and $C=0.39$. One can see that the
two fits are practically undistinguishable. Therefore the estimate
$\beta \simeq 0.01$~eV should be considered as upper limit of $\beta$
value. Thus its actual value can be much less than $0.01$~eV, though
still much larger than the average level spacing of the compound
nucleus~\cite{kun:97b}, for which the statistical model
calculations~\cite{blatt} yields $D\simeq 10^{-10}$~eV.

We recall that the total spin off-diagonal $S$-matrix correlations for
evaporation processes were justified in Ref.~\cite{kun:97b} in the
limit $N_{eff}\to\infty $, where $N_{eff}$ is an effective dimension
of the Hilbert space of the quasi-bound intermediate system. For the
analyzed Bi($\gamma$, p) photonuclear reaction we estimate
$N_{eff}\simeq \Gamma_{spr}/D \simeq 10^{16}$. Interestingly, the
condition of the exponentially large $N_{eff}$ for the anomalously
slow phase relaxation is consistent with the experimental data on a
proton thermal emission (evaporation) in proton induced nuclear
reactions~\cite{gugelot}. For heavy targets, Pt and Au, $N_{eff}\simeq
10^{20}$~\cite{flores} and the proton evaporating yields are strongly
forward peaked revealing that $\beta/\Gamma_{cn}\leq 1$. However, for
lighter targets Cu, Fe and Ni, $N_{eff}\simeq 10^{9}$~\cite{flores}
and the proton evaporating yields are symmetric about
90$^\circ$~c.m. indicating that $\beta/\Gamma_{cn}\gg 1$. Using the
statistical model formalism~\cite{blatt} we obtain that $\Gamma_{cn}$
for the Pt and Au targets is about 5 orders of magnitude smaller than
$\Gamma_{cn}$ for the Cu, Fe and Ni targets. Accordingly, the value of
$\beta$ for $N_{eff}\simeq 10^{16}$ is at least 6 orders of magnitude
or more smaller than the value of $\beta$ for $N_{eff}\simeq 10^{9}$.
The condition $N_{eff}\to\infty$ for the anomalously slow phase
relaxation also suggests that, with the decrease of the excitation
energy and the compound nucleus temperature,
$N_{eff}\simeq\Gamma_{spr}/D$ also decreases exponentially since $D$
increases exponentially while $\Gamma_{spr}$ is approximately
constant. Then, such a decrease of $N_{eff}$ is expected to lead to an
increase of the phase relaxation width,
$\beta\simeq\Gamma_{spr}\gg\Gamma_{cn}$. This means that, for
temperatures less than certain value, memory about the initial phase
relations is completely lost and the compound nucleus is no longer in
a superconducting state. This problem is worth an experimental study,
{\it e.g.}, for Pt(p,p$^\prime$) inelastic scattering~\cite{flores}.

There are many more data sets, including recent ones (see, {\it e.g.},
~\cite{flores}), demonstrating a strong, a factor two or more, forward
peaking for thermal emission in compound nucleus reactions. These
nuclear data will be analyzed in a future work. However, it should be
of interest to experimentally search for manifestations of thermalized
non-equilibrated matter in many-body mesoscopic systems and
nanostructures. For example, one may try to search for an asymmetry
around 90$^\circ$ in angular distributions of thermal electron yield
originated from the interaction of the electron beam with
many-electron quantum dots, often referred to as artificial nuclei.

It should be noted that while we have been able to determine an upper
limit of the anomalously small value of $\beta$ from the data
analysis, its theoretical evaluation is an open problem.

\section{Thermalized non-equilibrated matter as a high temperature
superconducting state}
\label{sec4}

Consider a proton beam directed on a heavy nucleus. Suppose the proton
from the incident beam is captured by the nucleus. As a result, a
thermalized compound nucleus, with strongly overlapping resonances,
$\Gamma_{cn}\gg D$,
 is formed. This compound nucleus can emit
a proton either with the energy lower than the energy of the incoming
proton or with the same energy as that of the incoming proton. The
latter possibility is referred to as compound elastic scattering. Since
the compound nucleus is formed due to the coherent contribution of
partial waves with orbital momenta ranging from $J=0$ to $J=J_{max}$,
only a fraction of the incoming plane wave contributes to the
formation of the thermalized compound nucleus. The intensity of this
fraction of the incoming plane wave is forward peaked with an angular
dispersion $\simeq 1/J_{max}$. Suppose that $J_{max}\beta
/\Gamma_{cn}\ll 1$, as it can be the case for the Pt(p,p') compound
nucleus scattering~\cite{MarcPRC06}. Then, the compound
elastic proton yield emitted from the fully thermalized (since
$\Gamma_{spr}\gg\Gamma_{cn}$~\cite{flores}) intermediate nucleus would
have precisely the same forward-peaked angular distribution, with the
same angular dispersion, as that for the fraction of the incoming
plane wave contributing to the formation of the compound nucleus. This
is because the phase relations between partial waves with different
angular momenta (total spins) for the emitted proton are the same as
for the incoming plane wave. In other words the incident beam passes
through the compound nucleus without any resistance. Such an ideal
transparency takes place in spite of a complete thermalization of the
intermediate compound nucleus having a high temperature. For this
reason, such a state of the thermalized compound nucleus can 
formally be referred to as a high temperature superconducting state in
strongly interacting quantum many-body system. It should be noted
that, on the basis of arguments in Ref.~\cite{kun:97b}, such a
superconducting state may be formed not only for the target nucleus
being in a ground state but also for highly-excited target nucleus.

On the contrary, for $J_{max}\beta /\Gamma_{cn}\gg 1$, the phase
relaxation-randomization is very fast, 
initial spin off-diagonal phase correlations
are completely forgotten. This results in the conventional Bohr's
picture of compound nucleus leading to symmetric angular distributions
around 90$^\circ$ c.m. Within this picture, half of the initial
forward-peaked incoming current would be emitted back implying an
infinite resistance for a proton wave propagation through the
thermalized compound nucleus: Equal forward and backward emission
intensities cancel each other resulting in zero net current.

\section{Conclusion}
\label{sec5}

The main purpose of this paper is not to promote the presented
description of the anomalously slow phase relaxation, which is many
orders of magnitude slower than energy relaxation
(thermalization). Rather, we intended to indicate a new field of
research in quantum many--body physics. Our intention is motivated by
an existence of nuclear data that reveal a clear physical picture for
a new form of matter -- thermalized non-equilibrated matter. The
problem is of importance not only for nuclear physics (and
applications for nuclear data evaluation) but it should be of interest
to a wider physics community. Indeed, the fact that in highly excited
many-body systems the phase relaxation can be many orders of magnitude 
longer than energy
relaxation is of significant implications for quantum
computing~\cite{flores,sigma} as well as, {\sl e.g.}, time-delayed
``statistical'' ionization of many-electron quantum dots and atomic
clusters (see, {\sl e.g.},~\cite{campb} and references therein).  A
clear presence of the effect of anomalously slow phase relaxation in
chemical reactions (see~\cite{BenetKW05,BCKW06} and references
therein) would require a modification of the statistical theories,
phase space and transition state theories (see, {\sl
  e.g.},~\cite{Levine} and references therein). And, as has been
discussed above, thermalized non-equilibrated matter may be viewed as
high temperature superconducting state in highly-excited 
quantum many-body systems.

Yet, the nuclear data indicating an existence of anomalously slow
phase relaxation, which is much slower than energy relaxation, have
been completely unrecognized by nuclear physicists and, for this
reason, are completely unknown outside the nuclear physics
community. In many fields, including statistical physics, the notions
of ``thermalization'' or ``energy equilibration'' are considered to be
equivalent to the notion ``statistical equilibrium''.

The conventional idea of a very fast phase relaxation in quantum
many-body systems is at the very foundation of the statistical model
and random matrix theory. Accordingly, it is widely presented in the
University courses on, {\sl e.g.,} nuclear physics, molecular and
atomic cluster physics, condensed matter, mesoscopic physics {\sl
  etc}. Yet, students should not be denied the opportunity and right
to know that other possibilities exist. Namely, that a thermalized
system is not necessarily in equilibrium due to the anomalously long
phase memory. These concepts may be counterintuitive, despite of the
sounding experimental evidence in their favor. Thus, a conceptual
revision of the long-standing conventional physical pictures is
required.

\begin{acknowledgments}
SK would like to thank the organizers of the First International
Meeting on Recent Developments in the Study of Radiation Effects in
Matter, Q. Roo, December 5-8, 2006 for a kind invitation to give a
talk and a warm hospitality.  One of us (SK) is grateful to Professor
Lew Chadderton for many useful discussions of the subject. This work
has been supported by the projects IN--111607 (DGAPA) and 43375
(CONACyT).
\end{acknowledgments}

\appendix

\section{Time-dependent correlation function of the many-particle
  density fluctuations for thermalized non-equilibrated matter}

All known superconductors are solids. None are gases or liquids. Does
our analysis of the nuclear data suggest a possibility of a high
temperature superconducting state for highly excited boiling ``nuclear
liquid''? In this Appendix we briefly outline some results of
Ref.~\cite{KUNunpubl} which indicate that this is not quite so.

The wave function of the compound nucleus may be written as 
\begin{equation}
\label{eqa1}
\Psi({\bf r},t)=\sum_{J\mu}\gamma_\mu^{J a}
\exp(-iE_\mu^{J}t/\hbar)\phi_\mu^J({\bf r}).
\end{equation}
Here, $\gamma_\mu^{J a}$ are real partial-width amplitudes for the
formation of the compound nucleus resonance level $\mu$ with the total
spin $J$ and energy $E_\mu^{J}$ from the channel $a$, $\phi_\mu^J({\bf
  r})$ are the resonance eigenstates with ${\bf r}$ denoting the
coordinates of all the particles, and $t$ is the time.  For each $J$
value the summation over $\mu$ includes $\simeq\Delta E/D\gg 1$ terms with
$\Delta E\simeq \Gamma_{spr}\gg \beta$, where $D$ is average level
spacing of the compound nucleus.  Summation over $J$-values goes from
$J=0$ to $J=J_{max}$.

We consider the correlation between the density fluctuations, $\delta
n({\bf r},t)$, for two different moments of time, $t_1$ and
$t_2$. Here,
\begin{equation}
\label{eqa2}
\delta n({\bf r},t)= n({\bf r},t)- {\overline{n({\bf r},t)}\,}^{\bf r}
\end{equation}
with $n({\bf r},t)=(1/V)|\Psi({\bf r},t)|^2$, $V$ being a
multi-dimensional effective volume of the system, and
${\overline{(...)}\,}^{\bf r}$ standing for $(1/V)\int_V d{\bf
  r}(...)$.  For $\hbar/D >t_1,t_2 >\hbar/\Delta E$ and $\beta/D\gg 1$,
we obtain~\cite{KUNunpubl}
\begin{eqnarray}
\label{eqa3}
{\overline{\delta n({\bf r},t_1)\delta n({\bf r},t_2)}\,}^{\bf r}& \simeq &
A\exp(-\Delta E|t_1-t_2|/\hbar)  \\
&  + & B\sum_{J_1J_2J_3J_4}\exp(-\beta|J_1-J_2|t_1/\hbar )
\exp(-\beta|J_3-J_4|t_2/\hbar )+R(t_1,t_2),  \nonumber
\end{eqnarray}
where the summation runs over all spin values except for $J_1=J_2$ and
$J_3=J_4$, and $A$ and $B$ are time--independent quantities. The first
term in the r.h.s.  of Eq.~(\ref{eqa3}) corresponds to a very quick
decay of the density correlations on the very short period of time
$|t_1-t_2|\simeq \hbar /\Delta E$. Yet, for $t_1=t_2$, this term does
not depend on $t_1$. This is characteristic of a liquid or gas
phase. On the contrary, in the second term the dependence on $t_1$ and
$t_2$ is factorized. Also, for $t_1,t_2\ll\hbar/(J_{max}\beta)$, the
second term does not depend on $t_1,t_2$ resembling the behaviour of a
solid (or glass).  Pictorially, this may be viewed as a certain, not
necessarily spherically symmetric, spatial configuration of ice put in
a liquid or gas. Notice that on this time scale,
$t_1,t_2\ll\hbar/(J_{max}\beta)$, the thermalized non-equilibrated
matter resembles a high temperature superconducting state.  As time
goes on the ``ice'' slowly melts down and, for $t
>\hbar/(J_{max}\beta)$, the density fluctuations due to the ``solid''
phase relaxes completely. This corresponds to the regime of complete
phase relaxation (complete loss of the phase memory) recovering the
conventional picture of the compound nucleus by Bohr. One also
observes that if only states with a single total spin value are
excited ($J_1=J_2=J_3=J_4$) the second term in the r.h.s. of
Eq.~(\ref{eqa3}) vanishes. Therefore, a formation of the ``solid''
phase is essentially due to the coherent excitation of and a peculiar
interference between the strongly overlapping ($t\ll \hbar/D$) states
with different $J$-values in Eq.~(\ref{eqa1}).

The third term in the r.h.s. of Eq.~(\ref{eqa3}), which vanishes for
$t_1,t_2>\hbar/(J_{max}\beta)$, corresponds to the interplay between
the ``solid'' phase and gas or liquid phase. This term will be
analyzed elsewhere.


\begin{thebibliography}{99}

\bibitem{campb}  
E.E.B. Campbell and R.D. Levine,
Ann. Rev. Phys. Chem. {\bf 51} 65 (2000).

\bibitem{flores} 
J. Flores, S.Yu. Kun, T.H. Seligman,
Phys. Rev. E {\bf 72} 017201 (2005); quant-ph/0502050.

\bibitem{sigma} 
M. Bienert, J. Flores, S.Yu. Kun, 
T.H. Seligman, Symmetry, Integrability and Geometry: Methods and
Applications (SIGMA) {\bf 2} paper 027 (2006); quant-ph/0602224.

\bibitem{kun97HI} 
S.Yu. Kun, Z. Phys. A {\bf 357} 271
(1997).

\bibitem{KunVagVor99} 
S.Yu. Kun, A.V. Vagov, O.K. Vorov,
 Phys. Rev. C {\bf 59} R585 (1999).

\bibitem{KunRobVag} 
S.Yu. Kun, B.A. Robson , A.V. Vagov,
Phys. Rev. Lett. {\bf 83} 504 (1999).

\bibitem{KunPRL00} 
S.Yu. Kun, Phys. Rev. Lett. {\bf 84} 423 (2000).

\bibitem{KVG01} 
S.Yu. Kun, A.V. Vagov, W. Greiner,
 Phys. Rev. C {\bf 63} 014608 (2001).

\bibitem{KunChVG02} 
S.Yu. Kun, L.T. Chadderton, A.V. Vagov,
W. Greiner, Int. J. Mod. Phys. E {\bf 11} 273 (2002).

\bibitem{KunBChGrH03} 
S.Yu. Kun, L. Benet,
L.T. Chadderton, W. Greiner, F. Haas, Phys. Rev. C {\bf 67} 011604(R) (2003);
 quant-ph/0205036.

\bibitem{BKWD05} L. Benet, S.Yu. Kun, Wang Qi, V. Denisov,
 Phys. Lett. B {\bf 605} 101 (2005); nucl-th/0407029.

\bibitem{BenetKW05} L. Benet, S.Yu. Kun, Wang Qi,
 Phys. Rev. C {\bf 73} 064602 (2006);
 quant-ph/0503046.

\bibitem{BCKW06} L. Benet, L.T. Chadderton, S.Yu. Kun, Wang Qi,
  ``Quantum-classical transition for an analog of double-slit
  experiment in complex collisions: Dynamical decoherence in quantum
  many-body systems'', quant-ph/0610091; submitted.

\bibitem{kun:94} S.Yu. Kun, Z. Phys. A {\bf 348} 273
(1994).

\bibitem{kun:97b} S.Yu. Kun, Z. Phys. A {\bf 357} 255
(1997).

\bibitem{Bigiantres64} R.R. Harvey, J.T. Caldwell,
R.L. Bramblett, S.C. Fultz, Phys. Rev. {\bf 136} B126
(1964). 

\bibitem{dataBi} M.E. Toms, W.E. Stephens, Phys. Rev. {\bf 92} 362 (1953).

\bibitem{blatt} J.M. Blatt and V.F. Weisskopf, {\sl Theoretical
Nuclear Physics} (Dover Publications, Inc., NY, 1991).

\bibitem{kun:93} S.Yu. Kun, Phys. Lett. B {\bf 319} 16
(1993).

\bibitem{ericson} T. Ericson and T. Mayer-Kuckuk,
Ann. Rev. Nucl. Sci. {\bf 16} 183 (1966).

\bibitem{AWM75} D. Agassi, H.A. Weidenm\"uller,
G. Mantzouranis, Phys. Rep. {\bf 22} 145 (1975). 

\bibitem{anderson} P.W. Anderson, {\sl Basic Notions of
Condensed Matter Physics} Frontiers in Physics, vol. 55, (The
Benjamin-Cummings, 1984) pp. 71--72.

\bibitem{gugelot} P.C. Gugelot, Phys. Rev. {\bf 93} 425
(1954).

\bibitem{MarcPRC06} M. Bienert, J. Flores, S.Yu. Kun,
 Phys. Rev. C {\bf 74} 027602 (2006); nucl-ex/0508020.

\bibitem{Levine} R.D. Levine and R.B. Bernstein, {\sl
Molecular Reaction Dynamics and Chemical Reactivity} (Oxford
Univ. Press, NY, 1987).

\bibitem{KUNunpubl} S.Yu. Kun, unpublished.

\end{thebibliography}
\end{document}